\documentclass[prb,twocolumn,showpacs]{revtex4}
\usepackage{graphicx} 

\usepackage{amsmath,amsthm,amssymb,mathrsfs}
\usepackage{bm}

\begin{document}

\title{Nonlinear analysis of magnetization dynamics excited by spin Hall effect}

\author{Tomohiro Taniguchi
      }
 \affiliation{
 National Institute of Advanced Industrial Science and Technology (AIST), Spintronics Research Center, Tsukuba, Ibaraki 305-8568, Japan. 
 }

\date{\today} 
\begin{abstract}
{
We investigate the possibility of exciting self-oscillation in a perpendicular ferromagnet by the spin Hall effect 
on the basis of a nonlinear analysis of the Landau-Lifshitz-Gilbert (LLG) equation. 
In the self-oscillation state, 
the energy supplied by the spin torque during a precession on a constant energy curve 
should equal the dissipation due to damping. 
Also, the current to balance the spin torque and the damping torque in 
the self-oscillation state should be larger than 
the critical current to destabilize the initial state. 
We find that these conditions in the spin Hall system are not satisfied 
by deriving analytical solutions of the energy supplied by the spin transfer effect 
and the dissipation due to the damping from the nonlinear LLG equation. 
This indicates that the self-oscillation of a perpendicular ferromagnet cannot be excited 
solely by the spin Hall torque. 
}
\end{abstract}

 \maketitle


\section{Introduction}
\label{sec:Introduction}

Nonlinear dynamics such as fast switching and self-oscillation (limit cycle) 
has been a fascinating topic in physics \cite{wiggins03,bertotti09text}. 
Magnetization dynamics excited by the spin transfer effect \cite{slonczewski96,berger96} 
in a nanostructured ferromagnet 
\cite{katine00,kiselev03,rippard04,kubota05,mangin06,houssameddine07,zeng12,kubota13} 
provide fundamentally important examples of such nonlinear dynamics. 
The magnetization switching was first observed in Co/Cu metallic multilayer 
in 2000 \cite{katine00}. 
Three years later, self-oscillation was reported in a similar system \cite{kiselev03}. 
In these early experiments on the spin transfer effect, 
linear analysis 
was used 
to estimate, for example, the critical current destabilizing 
the magnetization in equilibrium \cite{sun00,grollier03}. 
However, recently it became clear that 
nonlinear analysis is necessary to quantitatively analyze the magnetization dynamics 
\cite{bertotti09text,bertotti04,bertotti05,bertotti06,apalkov05,dykman12,newhall13,taniguchi13PRB1,taniguchi13PRB2,taniguchi13,taniguchi14APEX,pinna13,pinna14}. 
For example, current density to excite self-oscillation can be evaluated 
by solving a nonlinear vector equation called 
the Landau-Lifshitz-Gilbert (LLG) equation \cite{taniguchi13,taniguchi14APEX}. 


Originally, the spin transfer effect was studied by applying an electric current 
directly to a ferromagnetic multilayer. 
Recently, however, an alternative method employing the spin Hall effect has been used 
to observe the spin transfer effect 
\cite{ando08,miron10,miron11,liu12,liu12a,liu12b,niimi12,haney13,haney13a,kim13,kim14,torrejon14,tserkovnyak14,yu14}. 
The spin-orbit interaction in a nonmagnetic heavy metal scatters 
the spin-up and spin-down electrons to the opposite directions, 
producing a pure spin current flowing in the direction 
perpendicular to an applied current. 
The pure spin current excites the spin torque, called spin Hall torque, 
on a magnetization in a ferromagnet attached to a nonmagnet. 
The direction of the spin Hall torque is geometrically determined \cite{ando08}, 
and its magnitude shows a different angular dependence 
than the spin torque in the ferromagnetic multilayer \cite{slonczewski96}. 
Therefore, it is fundamentally unclear whether the physical phenomena observed in the multilayer 
\cite{katine00,kiselev03,rippard04,kubota05,mangin06,houssameddine07,zeng12,kubota13} 
can be reproduced in the spin Hall system, 
and thus, new physical analysis is necessary. 
The magnetization switching of both in-plane magnetized 
and perpendicularly magnetized ferromagnets by spin Hall torque 
was recently reported \cite{miron10,miron11,liu12,liu12a,kim13,kim14}. 
Accordingly, it might be reasonable to expect reports on 
self-oscillation by spin Hall torque. 
However, whereas self-oscillation has been observed in the in-plane magnetized system \cite{liu12b}, 
it has not been reported yet in the perpendicularly magnetized system. 


The purpose of this paper is to investigate 
the possibility of exciting self-oscillation by spin Hall torque 
based on a nonlinear analysis of the LLG equation. 
We argue that two physical conditions should be satisfied to excite self-oscillation. 
The first condition is that 
the energy that the spin torque supplies during a precession on a constant energy curve 
should equal the dissipation due to damping. 
The second condition is that 
the current to balance the spin torque and the damping torque in 
the self-oscillation state should be larger than 
the critical current to destabilize the initial state. 
This is because the magnetization initially stays at the minimum energy state, 
whereas the self-oscillation corresponds to a higher energy state. 
We derive exact solutions of the energy supplied by the spin transfer effect 
and the dissipation due to damping 
in the spin Hall system by solving the nonlinear LLG equation, 
and find that these conditions are not satisfied. 
Thus, the self-oscillation of a perpendicular ferromagnet cannot be excited 
solely by the spin Hall torque. 


The paper is organized as follows. 
The physical conditions to excite a self-oscillation is summarized in Sec. \ref{sec:Physical conditions to excite self-oscillation}. 
These conditions are applied to the spin Hall system in Sec. \ref{sec:Spin Hall system}. 
Section \ref{sec:Conclusion} is devoted to the conclusions. 


\section{Physical conditions to excite self-oscillation}
\label{sec:Physical conditions to excite self-oscillation}

Let us first summarize the physical conditions necessary to excite self-oscillation. 
The magnetization dynamics are described by the LLG equation 
\begin{equation}
  \frac{d \mathbf{m}}{dt}
  =
  -\gamma
  \mathbf{m}
  \times
  \mathbf{H}
  -
  \gamma
  H_{\rm s}
  \mathbf{m}
  \times
  \left(
    \mathbf{p}
    \times
    \mathbf{m}
  \right)
  +
  \alpha
  \mathbf{m}
  \times
  \frac{d\mathbf{m}}{dt},
  \label{eq:LLG}
\end{equation}
where $\mathbf{m}$ and $\mathbf{p}$ are 
the unit vectors pointing in the directions of 
the magnetization and the spin polarization of the spin current, respectively. 
The gyromagnetic ratio and the Gilbert damping constant are denoted as $\gamma$ and $\alpha$, respectively. 
The magnetic field $\mathbf{H}$ relates to the energy density of the ferromagnet $E$ via 
$\mathbf{H}=-\partial E/\partial (M \mathbf{m})$, 
where $M$ is the saturation magnetization. 
The strength of the spin torque, $H_{\rm s}$, is proportional to the current density $j$. 
Since the LLG equation conserves the norm of the magnetization, 
the magnetization dynamics can be described as a trajectory on a unit sphere. 
The energy density $E$ shows constant energy curves on this sphere. 
For example, when the system has uniaxial anisotropy, 
the constant energy curves are latitude lines. 
The self-oscillation is a steady precession state on a constant energy curve 
excited by the field torque, the first term on the right-hand side of Eq. (\ref{eq:LLG}). 
This means that the second and third terms 
of Eq. (\ref{eq:LLG}), averaged over the constant energy curve, cancel each other. 
In other words, the energy supplied by the spin transfer effect during the precession on the constant energy curve 
equals the dissipation due to the damping. 
This condition can be expressed as \cite{bertotti09text,taniguchi14APEX}
\begin{equation}
  \oint 
  dt 
  \frac{dE}{dt}
  =
  \mathscr{W}_{\rm s}
  +
  \mathscr{W}_{\alpha}
  =
  0,
  \label{eq:condition}
\end{equation}
where the energy supplied by the spin transfer effect 
and the dissipation due to the damping 
during the precession on the constant energy curve of $E$ are given by 
\cite{bertotti09text,bertotti04,bertotti05,bertotti06,apalkov05,dykman12,newhall13,taniguchi13PRB1,taniguchi13PRB2,taniguchi13,taniguchi14APEX,pinna13,pinna14}
\begin{equation}
  \mathscr{W}_{\rm s}(E)
  =
  \gamma 
  M 
  \oint 
  dt 
  H_{\rm s}
  \left[
    \mathbf{p}
    \cdot
    \mathbf{H}
    -
    \left(
      \mathbf{m}
      \cdot
      \mathbf{p}
    \right)
    \left(
      \mathbf{m}
      \cdot
      \mathbf{H}
    \right)
  \right],
  \label{eq:Melnikov_s}
\end{equation}
\begin{equation}
  \mathscr{W}_{\alpha}(E)
  =
  -\alpha 
  \gamma 
  M 
  \oint 
  dt 
  \left[
    \mathbf{H}^{2}
    -
    \left(
      \mathbf{m}
      \cdot
      \mathbf{H}
    \right)^{2}
  \right].
  \label{eq:Melnikov_alpha}
\end{equation}
The time integral is over a precession period 
on a constant energy curve. 
We emphasize that Eqs. (\ref{eq:Melnikov_s}) and (\ref{eq:Melnikov_alpha}) are 
functions of the energy density $E$. 
We denote the minimum and maximum values of $E$ as $E_{\rm min}$ and $E_{\rm max}$, respectively. 
When the energy density also has saddle points $E_{\rm saddle}$, 
$E_{\rm max}$ in the following discussion can be replaced by $E_{\rm saddle}$. 
To excite the self-oscillation, 
there should be a certain value of the electric current density 
that satisfies Eq. (\ref{eq:condition}) for $E_{\rm min} < E < E_{\rm max}$ 
in a set of real numbers. 
Therefore, Eq. (\ref{eq:condition}) can be rewritten as 
\begin{equation}
  \exists
  j
  \in
  \mathbb{R},\ 
  \mathscr{W}_{\rm s}
  +
  \mathscr{W}_{\alpha}
  =
  0.
  \label{eq:condition_1}
\end{equation}
We denote the current satisfying the first condition, 
Eq. (\ref{eq:condition}), or equivalently Eq. (\ref{eq:condition_1}), as $j(E)$. 


Another condition necessary to excite self-oscillation 
relates to the fact that 
the magnetization initially stays at the minimum energy state. 
To excite any kind of magnetization dynamics, 
the spin torque should destabilize the initial state, 
which means that 
a current density larger than the critical current density, $j_{\rm c}=j(E_{\rm min})$, should be injected. 
Then, the condition 
\begin{equation}
  j(E) 
  > 
  j(E_{\rm min}),
  \label{eq:condition_2}
\end{equation}
should be satisfied to excite the self-oscillation. 
If this condition is not satisfied, 
the magnetization directly moves to a constant energy curve including the saddle point 
without showing a stable steady precession, and stops dynamics 
because the spin torque does not balance the damping torque for $E_{\rm min} < E < E_{\rm saddle}$. 
An example of such dynamics is shown below; see Fig. \ref{fig:fig3}. 
We emphasize that Eqs. (\ref{eq:condition_1}) and (\ref{eq:condition_2}) are applicable to any kind of physical system 
showing a self-oscillation. 



\begin{figure}
\centerline{\includegraphics[width=0.65\columnwidth]{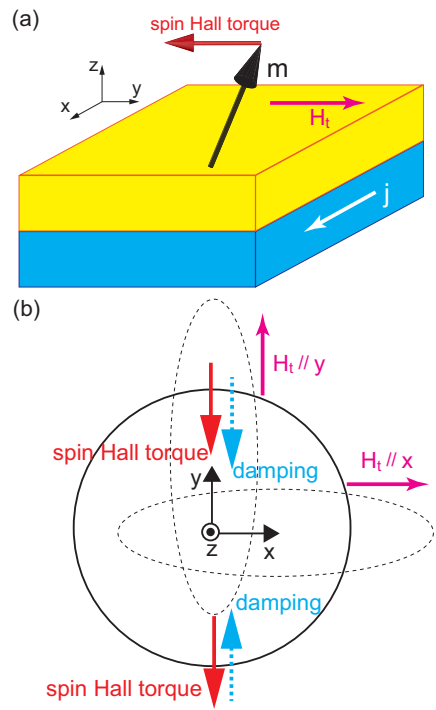}}
\caption{
         (a) Schematic view of system. 
             The current density $j$ flows in the nonmagnet along the $x$-axis, 
             exciting the spin Hall torque pointing in the $y$-direction on the magnetization $\mathbf{m}$ in the ferromagnet. 
             The applied magnetic field is denoted as $H_{\rm t}$. 
         (b) Schematic view of the precession trajectory of the magnetization on the constant energy curve. 
             The solid circle is the trajectory in the absence of the magnetic field or in the presence of the field along the $z$-axis, 
             whereas the dashed elliptical lines are those in the presence of the field in the $x$ and $y$-axes. 
             The solid and dotted arrows represent the directions of the spin Hall torque and the damping torque, respectively. 
         \vspace{-3ex}}
\label{fig:fig1}
\end{figure}





\section{Spin Hall system}
\label{sec:Spin Hall system}

Let us apply the above discussions to the spin Hall system 
schematically shown in Fig. \ref{fig:fig1} (a), 
where the electric current flows in the nonmagnet along the $x$ direction, 
whereas the ferromagnet is attached along the $z$ direction. 
The spin polarization of the spin current is geometrically determined as $\mathbf{p}=\mathbf{e}_{y}$. 
In the spin Hall system, 
the spin torque strength $H_{\rm s}$ is given by 
\begin{equation}
  H_{\rm s}
  =
  \frac{\hbar \vartheta j}{2eMd},
  \label{eq:H_s}
\end{equation}
where $\vartheta$ and $d$ are the spin Hall angle 
and the thickness of the ferromagnet, respectively. 
The magnetic field $\mathbf{H}$ consists of 
the applied field $H_{\rm t}$ and 
the perpendicular anisotropy field $H_{\rm K}m_{z} \mathbf{e}_{z}$. 
We can assume that $H_{\rm t}>0$ without losing generality 
because the sign of $H_{\rm t}$ only affects the sign of $j(E)$ derived below. 
Since we are interested in a perpendicular ferromagnet, 
we assume that $H_{\rm K}>H_{\rm t}>0$. 
Figure \ref{fig:fig1} (b) schematically shows 
the precession trajectory of the magnetization on a constant energy curve, 
where the directions of the spin Hall torque and the damping torque are 
represented by the solid and dotted arrows, respectively. 
The spin Hall torque is parallel to the damping torque for $m_{y}>0$, 
whereas it is anti-parallel to the damping torque for $m_{y}<0$. 
This means that the spin Hall torque dissipates energy from the ferromagnet when $m_{y}>0$, 
and supplies the energy to the ferromagnet when $m_{y}<0$. 
Then, due to the symmetry of the trajectory, 
the net energy supplied by the spin Hall torque, $\mathscr{W}_{\rm s}$, is zero 
when the applied magnetic field points to the $x$- or $z$-direction. 
This means that Eq. (\ref{eq:condition}) cannot be satisfied, 
and thus, self-oscillation cannot be excited in the spin Hall system 
in the absence of the applied magnetic field, 
or in the presence of the field pointing in the $x$- or $z$-direction. 
Therefore, in the following 
we focus on the applied magnetic field pointing in the $y$-direction. 
The magnetic field and the energy density are given by 
\begin{equation}
  \mathbf{H}
  =
  H_{\rm t}
  \mathbf{e}_{y}
  +
  H_{\rm K}
  m_{z}
  \mathbf{e}_{z},
\end{equation}
\begin{equation}
  E
  =
  -MH_{\rm t}
  m_{y}
  -
  \frac{MH_{\rm K}}{2}
  m_{z}^{2}.
  \label{eq:energy}
\end{equation}
The minimum energy of Eq. (\ref{eq:energy}) is 
\begin{equation}
  E_{\rm min}
  =
  -\frac{MH_{\rm K}}{2}
  \left[
    1
    +
    \left(
      \frac{H_{\rm t}}{H_{\rm K}}
    \right)^{2}
  \right],
  \label{eq:ene_min}
\end{equation}
which corresponds to a point $\mathbf{m}_{\rm stable}=(0,H_{\rm t}/H_{\rm K},\sqrt{1-(H_{\rm t}/H_{\rm K})^{2}})$. 
On the other hand, Eq. (\ref{eq:energy}) has a saddle point at $\mathbf{m}_{\rm saddle}=(0,1,0)$, 
corresponding to the energy density 
\begin{equation}
  E_{\rm saddle}
  =
  -MH_{\rm t}. 
  \label{eq:ene_saddle}
\end{equation}
Since the magnetization initially stays at the minimum energy state, 
and the magnetization dynamics stops when $\mathbf{m}$ reaches the saddle point $\mathbf{m}_{\rm saddle}$, 
we consider the energy region of $E_{\rm min} < E < E_{\rm saddle}$. 
To calculate Eqs. (\ref{eq:Melnikov_s}) and (\ref{eq:Melnikov_alpha}), 
it is necessary to solve a nonlinear equation $d \mathbf{m}/dt=-\gamma \mathbf{m} \times \mathbf{H}$, 
which determines the precession trajectory of $\mathbf{m}$ on the constant energy curve. 
Since the constant energy curve of Eq. (\ref{eq:energy}) is symmetric with respect to the $yz$-plane, 
it is sufficient for the calculation of 
Eqs. (\ref{eq:Melnikov_s}) and (\ref{eq:Melnikov_alpha}) 
to derive the solutions of $\mathbf{m}$ for half of the trajectory 
in the region of $m_{x}>0$, 
which are exactly given by 
\begin{equation}
  m_{x}(E)
  =
  (r_{2}-r_{3})
  {\rm sn}(u,k)
  {\rm cn}(u,k),
  \label{eq:mx_E}
\end{equation}
\begin{equation}
  m_{y}(E)
  =
  r_{3}
  +
  (r_{2}-r_{3})
  {\rm sn}^{2}(u,k),
  \label{eq:my_E}
\end{equation}
\begin{equation}
  m_{z}(E)
  =
  \sqrt{
    1
    -
    r_{3}^{2}
    -
    (r_{2}^{2}-r_{3}^{2})
    {\rm sn}^{2}(u,k)
  },
  \label{eq:mz_E}
\end{equation}
where $u=\gamma \sqrt{H_{\rm t} H_{\rm K}/2}\sqrt{r_{1}-r_{3}}t$, and 
$r_{\ell}$ are given by 
\begin{equation}
  r_{1}(E)
  =
  -\frac{E}{MH_{\rm t}},
  \label{eq:r_1}
\end{equation}
\begin{equation}
  r_{2}(E)
  =
  \frac{H_{\rm t}}{H_{\rm K}}
  +
  \sqrt{
    1
    +
    \left(
      \frac{H_{\rm t}}{H_{\rm K}}
    \right)^{2}
    +
    \frac{2E}{MH_{\rm K}}
  },
  \label{eq:r_2}
\end{equation}
\begin{equation}
  r_{3}(E)
  =
  \frac{H_{\rm t}}{H_{\rm K}}
  -
  \sqrt{
    1
    +
    \left(
      \frac{H_{\rm t}}{H_{\rm K}}
    \right)^{2}
    +
    \frac{2E}{MH_{\rm K}}
  }.
  \label{eq:r_3}
\end{equation}
The modulus of Jacobi elliptic functions, ${\rm sn}(u,k)$ and ${\rm cn}(u,k)$, is 
\begin{equation}
  k
  =
  \sqrt{
    \frac{r_{2}-r_{3}}{r_{1}-r_{3}}
  }.
  \label{eq:modulus}
\end{equation}
The derivations of Eqs. (\ref{eq:mx_E}), (\ref{eq:my_E}), and (\ref{eq:mz_E}) are shown in Appendix A. 
The precession period is 
\begin{equation}
  \tau(E)
  =
  \frac{2 \mathsf{K}(k)}{\gamma \sqrt{H_{\rm t} H_{\rm K}/2} \sqrt{r_{1}-r_{3}}},
  \label{eq:period}
\end{equation}
where $\mathsf{K}(k)$ is the first kind of complete elliptic integral. 
The work done by spin torque and the dissipation due to damping, $\mathscr{W}_{\rm s}$ and $\mathscr{W}_{\alpha}$, are obtained 
by substituting Eqs. (\ref{eq:mx_E}), (\ref{eq:my_E}), and (\ref{eq:mz_E}) into Eqs. (\ref{eq:Melnikov_s}) and (\ref{eq:Melnikov_alpha}), 
integrating over $[0,\tau/2]$, 
and multiplying a numerical factor 2 
because Eqs. (\ref{eq:mx_E}), (\ref{eq:my_E}), and (\ref{eq:mz_E}) are the solution of the precession trajectory for a half period. 
Then, $\mathscr{W}_{\rm s}$ and $\mathscr{W}_{\alpha}$ 
for $E_{\rm min} < E < E_{\rm saddle}$ are exactly given by 
\begin{equation}
  \mathscr{W}_{\rm s}
  =
  \frac{8MH_{\rm s} \sqrt{r_{1}-r_{3}}}{3 H_{\rm t} \sqrt{H_{\rm K}/(2 H_{\rm t})}}
  \mathcal{H}_{\rm s},
  \label{eq:W_s}
\end{equation}
\begin{equation}
  \mathscr{W}_{\alpha}
  =
  -\frac{4\alpha M \sqrt{r_{1}-r_{3}}}{3 \sqrt{H_{\rm K}/(2 H_{\rm t})}}
  \mathcal{H}_{\alpha},
  \label{eq:W_alpha}
\end{equation}
where $\mathcal{H}_{\rm s}$ and $\mathcal{H}_{\alpha}$ are given by 
\begin{equation}
\begin{split}
  \mathcal{H}_{\rm s}
  =&
  H_{\rm t}
  \left(
    \frac{1-r_{1}^{2}}{r_{1}-r_{3}}
  \right)
  \mathsf{K}(k)
  -
  \left(
    \frac{E}{M}
    +
    \frac{H_{\rm t}^{2}}{H_{\rm K}}
  \right)
  \mathsf{E}(k),
  \label{eq:H_s}
\end{split}
\end{equation}
\begin{equation}
\begin{split}
  \mathcal{H}_{\alpha}
  =&
  H_{\rm t}
  \left(
    \frac{1-r_{1}^{2}}{r_{1}-r_{3}}
  \right)
  \mathsf{K}(k)
  +
  \left(
    \frac{5E}{M}
    +
    3 H_{\rm K}
    +
    \frac{2H_{\rm t}^{2}}{H_{\rm K}}
  \right)
  \mathsf{E}(k).
  \label{eq:H_alpha}
\end{split}
\end{equation}
Here, $\mathsf{E}(k)$ is the second kind of complete elliptic integral. 
The derivations of Eqs. (\ref{eq:W_s}) and (\ref{eq:W_alpha}) are shown in Appendix B. 
The current $j(E)$ for $E_{\rm min} < E < E_{\rm saddle}$ is given by 
\begin{equation}
  j(E)
  =
  \frac{2 \alpha eMd}{\hbar \vartheta}
  \frac{H_{\rm t} \mathcal{H}_{\alpha}}{2 \mathcal{H}_{\rm s}}.
  \label{eq:jE}
\end{equation}
The currents for $E \to E_{\rm min}$ and $E \to E_{\rm saddle}$ are \cite{comment1} 
\begin{equation}
  j(E_{\rm min})
  =
  \frac{2 \alpha eMd}{\hbar \vartheta}
  \frac{H_{\rm K}}{H_{\rm t}/H_{\rm K}}
  \left[
    1
    -
    \frac{1}{2}
    \left(
      \frac{H_{\rm t}}{H_{\rm K}}
    \right)^{2}
  \right],
  \label{eq:jE_min}
\end{equation}
\begin{equation}
  j(E_{\rm saddle})
  =
  \frac{2 \alpha eMd}{\hbar \vartheta}
  \left(
    \frac{3 H_{\rm K} - 2 H_{\rm t}}{2}
  \right).
  \label{eq:jE_saddle}
\end{equation}



\begin{figure}
\centerline{\includegraphics[width=0.9\columnwidth]{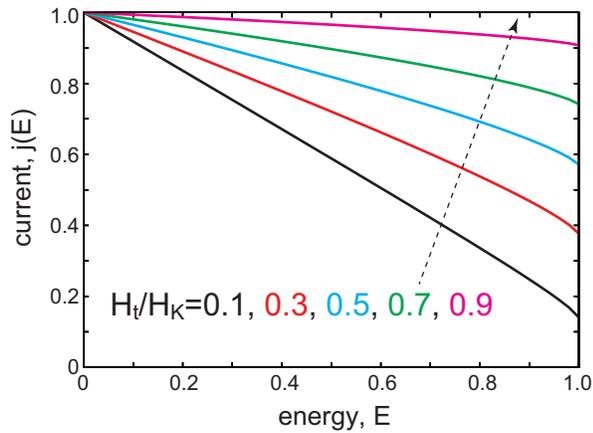}}
\caption{
         The dependence of the current $j(E)$, Eq. (\ref{eq:jE}), for several values of $H_{\rm t}/H_{\rm K}$ on the energy density $E$. 
         For simplicity, the horizontal and vertical axes are normalized as 
         $j(E)/j_{\rm c}$ and $E/(E_{\rm saddle}-E_{\rm min})-[E_{\rm min}/(E_{\rm saddle}-E_{\rm min})]$ 
         to make $j(E_{\rm min})=1$, $E_{\rm min}=0$, and $E_{\rm saddle}=1$. 
         \vspace{-3ex}}
\label{fig:fig2}
\end{figure}





\begin{figure}
\centerline{\includegraphics[width=0.9\columnwidth]{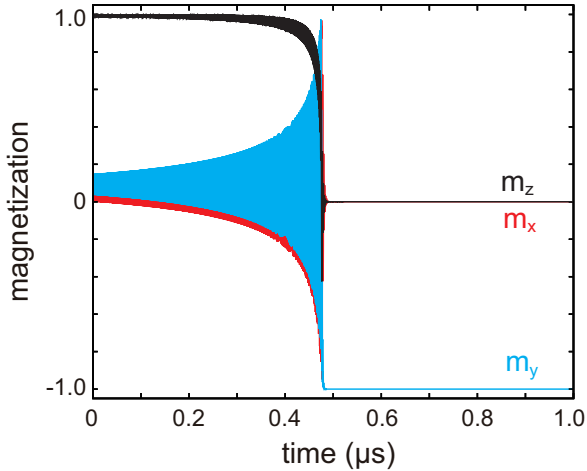}}
\caption{
         Typical magnetization dynamics excited by the spin Hall effect. 
         The parameter values are taken from experiments \cite{kim13,kim14,torrejon14,comment_Hayashi} as 
         $M=1500$ emu/c.c., $H_{\rm K}=540$ Oe, $\alpha=0.005$, $\gamma=1.764 \times 10^{7}$ rad/(Oe$\cdot$s), $d=1$ nm, $\vartheta=0.1$, and $H_{\rm t}=50$ Oe. 
         The current magnitude is $14 \times 10^{6}$ A/cm${}^{2}$, 
         while the critical current, Eq. (\ref{eq:jE_min}), is $13 \times 10^{6}$ A/cm${}^{2}$. 
         \vspace{-3ex}}
\label{fig:fig3}
\end{figure}




Equation (\ref{eq:jE}) is the current density 
satisfying Eq. (\ref{eq:condition}), or equivalently Eq. (\ref{eq:condition_1}). 
Then, let us investigate whether Eq. (\ref{eq:jE}) satisfies Eq. (\ref{eq:condition_2}). 
It is mathematically difficult to calculate 
the derivative of Eq. (\ref{eq:jE}) with respect to $E$ for an arbitrary value of $E$, 
although we can confirm that $j(E_{\rm min}) > j(E_{\rm saddle})$ for $H_{\rm t}<H_{\rm K}$. 
We note that a parameter determining whether Eq. (\ref{eq:condition_2}) is satisfied is 
only $H_{\rm t}/H_{\rm K}$ because the other parameters, such as $\alpha$ and $M$, are just 
common prefactors for any $j(E)$. 
As shown in Fig. \ref{fig:fig2}, 
$j(E)$ is a monotonically decreasing function of $E$ for a wide range of $H_{\rm t}/H_{\rm K}$, 
i.e., Eq. (\ref{eq:condition_2}) is not satisfied. 
This result indicates that 
the magnetization stays in the equilibrium state when $j<j_{\rm c}=j(E_{\rm min})$, 
whereas it moves to the constant energy curve of $E_{\rm saddle}$ 
without showing stable self-oscillation when $j>j_{\rm c}$ 
because the spin Hall torque does not balance the damping torque 
on any constant energy curve between $E_{\rm min}$ and $E_{\rm saddle}$. 
The magnetization finally stops its dynamics at $\pm\mathbf{m}_{\rm saddle}$ 
because all torques become zero at these points. 
Figure \ref{fig:fig3} shows a typical example of such dynamics, 
in which the time evolution of each component is shown. 
Therefore, self-oscillation solely by the spin Hall torque 
cannot be excited in the perpendicular ferromagnet. 
This is a possible reason why the self-oscillation has not been reported yet. 


Recently, many kinds of other torques pointing in different directions 
or having different angular dependencies, 
such as field-like and Rashba torques, have been proposed 
\cite{miron10,miron11,kim13,kim14,yu14,pauyac13,garello13,qiu14}. 
These effects might change the above conclusions. 
Adding an in-plane anisotropy \cite{taniguchi13PRB1,taniguchi13PRB2}, 
tilting the perpendicular anisotropy \cite{yu14}, 
or using higher order anisotropy 
might be another candidate. 
Spin pumping is also an interesting phenomenon because it modifies the Gilbert damping constant 
\cite{tserkovnyak02a,tserkovnyak02b,tserkovnyak03,ciccarelli15}. 
It was shown in Refs. \cite{tserkovnyak03,taniguchi07} that 
the enhancement of the Gilbert damping constant 
in a ferromagnetic/nonmagnetic/ferromagnetic trilayer system depends on 
the relative angle of the magnetization. 
This means that the Gilbert damping constant has an angular dependence. 
In a such case, it might be possible to satisfy Eqs. (\ref{eq:condition_1}) and (\ref{eq:condition_2}) 
by attaching another ferromagnet to the spin Hall system 
and by choosing an appropriate alignment of the magnetizations. 
The above formulas also apply to these studies. 
In Appendix C, we briefly discuss a technical difficulty to include the effect of 
the field-like torque or Rashba torque. 





\section{Conclusion}
\label{sec:Conclusion}

In conclusion, 
we developed a method for the nonlinear analysis of the LLG equation 
in the spin Hall system with a perpendicular ferromagnet. 
We summarized physical conditions to excite 
self-oscillation by the spin transfer effect. 
The first condition, Eq. (\ref{eq:condition}), or equivalently Eq. (\ref{eq:condition_1}), implies that 
the energy supplied by the spin torque during a precession on a constant energy curve 
should equal the dissipation due to damping. 
The second condition, Eq. (\ref{eq:condition_2}), implies that 
the current to balance the spin torque and the damping torque in 
the self-oscillation state should be larger than 
the critical current to destabilize the initial state. 
By solving the nonlinear LLG equation, 
we derived exact solutions of the energy supplied by the spin transfer effect 
and the dissipation due to damping, 
and showed that these conditions are not satisfied. 
These results indicate that self-oscillation cannot be excited solely by the spin Hall torque. 


The author would like to acknowledge T. Yorozu for his great constructive help on this work. 
The author also thanks M. Hayashi, H. Kubota, and A. Emura for their kind supports. 
This work was supported by JSPS KAKENHI Grant-in-Aid for Young Scientists (B) 25790044. 


\appendix


\section{Precession trajectory on a constant energy curve}

Here, we show the derivation of Eqs. (\ref{eq:mx_E}), (\ref{eq:my_E}), and (\ref{eq:mz_E}). 
The precession trajectory on a constant energy curve is determined by 
$d \mathbf{m}/dt = -\gamma \mathbf{m} \times \mathbf{H}$. 
The $y$-component of this equation is $d m_{y}/dt = \gamma H_{\rm K} m_{x} m_{z}$. 
Thus, we find 
\begin{equation}
  \int 
  dt 
  =
  \frac{1}{\gamma H_{\rm K}}
  \int 
  \frac{d m_{y}}{m_{x} m_{z}}. 
  \label{eq:time_integral}
\end{equation}
As mentioned in Sec. \ref{sec:Spin Hall system}, 
since the constant energy curve of Eq. (\ref{eq:energy}) is symmetric with respect to the $yz$-plane, 
it is sufficient to derive the solutions of $\mathbf{m}$ for half of the trajectory 
in the region of $m_{x}>0$. 
Using $E$ and $m_{y}$, 
$m_{x}$ and $m_{z}$ are expressed as 
\begin{equation}
  m_{x}
  =
  \sqrt{
    1
    -
    m_{y}^{2}
    +
    \frac{2E}{MH_{\rm K}}
    +
    \frac{2H_{\rm t}}{H_{\rm K}}
    m_{y}
  },
  \label{eq:mx_E_my}
\end{equation}
\begin{equation}
  m_{z}
  =
  \sqrt{
    -\frac{2E}{MH_{\rm K}}
    -
    \frac{2 H_{\rm t}}{H_{\rm K}}
    m_{y}
  }.
  \label{eq:mz_E_my}
\end{equation}
The initial state of $m_{y}$ is chosen as $m_{y}(0)=r_{3}$, 
where $r_{3}$ is given by Eq. (\ref{eq:r_3}). 
Then, $m_{y}$ at a certain time $t$ is determined from 
Eq. (\ref{eq:time_integral}) as 
\begin{equation}
\begin{split}
&
  \gamma
  \sqrt{2 H_{\rm t} H_{\rm K}}
  \int_{0}^{t} dt 
\\
  &=
  \int_{r_{3}}^{m_{y}} 
  \frac{d m_{y}^{\prime}}{\sqrt{ (m_{y}^{\prime}-r_{1}) (m_{y}^{\prime} - r_{2}) (m_{y}^{\prime}-r_{3})}}.
\end{split}
\end{equation}
We introduce a new parameter $s$ as $m_{y}=r_{3}+(r_{2}-r_{3})s^{2}$. 
Then, we find 
\begin{equation}
  \gamma
  \sqrt{
    \frac{H_{\rm t} H_{\rm K}}{2}
  }
  \sqrt{r_{1}-r_{3}}
  t
  =
  \int_{0}^{s} 
  \frac{d s^{\prime}}{\sqrt{(1-s^{\prime 2}) ( 1-k^{2}s^{\prime 2})}},
\end{equation}
where the modulus $k$ is given by Eq. (\ref{eq:modulus}). 
The solution of $s$ is $s={\rm sn}(u,k)$. 
Therefore, $m_{y}$ is given by Eq. (\ref{eq:my_E}). 
Equations (\ref{eq:mx_E}) and (\ref{eq:mz_E}) are obtained 
by substituting Eq. (\ref{eq:my_E}) into Eqs. (\ref{eq:mx_E_my}) and (\ref{eq:mz_E_my}). 


We note that Eqs. (\ref{eq:mx_E}), (\ref{eq:my_E}), and (\ref{eq:mz_E}) are periodic functions 
with the period given by  Eq. (\ref{eq:period}). 
On the other hand, when $E=E_{\rm saddle}$, 
the magnetization stops its dynamics finally at the saddle point $\mathbf{m}=(0,1,0)$. 
The solution of the constant energy curve of $E_{\rm saddle}$ with the initial condition $m_{y}(0)=r_{3}$ 
can be obtained by similar calculations, 
and are given by 
\begin{equation}
  m_{x}
  =
  2 
  \left(
    1
    -
    \frac{H_{\rm t}}{H_{\rm K}}
  \right)
  \frac{\tanh(\nu t)}{\cosh(\nu t)},
  \label{eq:mx_saddle}
\end{equation}
\begin{equation}
  m_{y}
  =
  -1
  +
  \frac{2H_{\rm t}}{H_{\rm K}}
  +
  2
  \left(
    1
    -
    \frac{H_{\rm t}}{H_{\rm K}}
  \right)
  \tanh^{2}(\nu t),
  \label{eq:my_saddle}
\end{equation}
\begin{equation}
  m_{z}
  =
  2 
  \sqrt{
    \frac{H_{\rm t}}{H_{\rm K}}
    \left(
      1
      -
      \frac{H_{\rm t}}{H_{\rm K}}
    \right)
  }
  \frac{1}{\cosh(\nu t)},
  \label{eq:mz_saddle}
\end{equation}
where $\nu=\gamma \sqrt{H_{\rm t}(H_{\rm K}-H_{\rm t})}$. 


\section{Derivation of Eqs. (\ref{eq:W_s}) and (\ref{eq:W_alpha})}

Using Eqs. (\ref{eq:mx_E}), (\ref{eq:my_E}), and (\ref{eq:mz_E}), 
the explicit form of Eq. (\ref{eq:Melnikov_s}) for the spin Hall system is given by 
$\mathscr{W}_{\rm s}=\gamma M H_{\rm s} \int dt w_{\rm s}$, 
where $w_{\rm s}$ is given by 
\begin{equation}
\begin{split}
  &
  w_{\rm s}
  =
  \left(
    H_{\rm t}
    -
    H_{\rm K}
    r_{3}
  \right)
  (1-r_{3}^{2})
\\
  &+
  \left\{
    -2H_{\rm t}
    r_{3}
    +
    H_{\rm K}
    \left[
      r_{3}
      (r_{2}+r_{3})
      -
      (1-r_{3}^{2})
    \right]
  \right\}
  (r_{2}-r_{3})
  {\rm sn}^{2}(u,k)
\\
  &+
  \left\{
    -H_{\rm t}
    +
    H_{\rm K}
    (r_{2}+r_{3})
  \right\}
  (r_{2}-r_{3})^{2}
  {\rm sn}^{4}(u,k). 
  \label{eq:w_s_tmp}
\end{split}
\end{equation}
Similarly, Eq. (\ref{eq:W_alpha}) for the spin Hall system is given by 
$\mathscr{W}_{\alpha}=-\alpha \gamma M \int dt w_{\alpha}$, 
where $w_{\alpha}$ is given by 
\begin{equation}
\begin{split}
&
  w_{\alpha}
  =
  (1-r_{3}^{2})
  (H_{\rm t}-H_{\rm K}r_{3})^{2}
\\
  &-
  \left[
    2 H_{\rm t}^{2}
    r_{3}
    -
    H_{\rm K}^{2}
    (r_{2}+r_{3})
    (1 - 2 r_{3}^{2})
    +
    2 H_{\rm t}
    H_{\rm K}
    (1-r_{2}r_{3}-2r_{3}^{2})
  \right]
\\
  &\ \ \ \ \ \times
  (r_{2}-r_{3})
  {\rm sn}^{2}(u,k)
\\
  &-
  \left[
    H_{\rm t}
    -
    H_{\rm K} (r_{2}+r_{3})
  \right]^{2}
  (r_{2}-r_{3})^{2}
  {\rm sn}^{4}(u,k). 
\end{split}
\end{equation}
Then, $\mathscr{W}_{\rm s}$ and $\mathscr{W}_{\alpha}$ are obtained 
by integrating over $[0,\tau/2]$, 
and multiplying a numerical factor 2. 
The following integral formulas are useful, 
\begin{equation}
  \int^{u} 
  du^{\prime} 
  {\rm sn}^{2}(u^{\prime},k)
  =
  \frac{u - \mathsf{E}[{\rm am}(u,k),k]}{k^{2}},
\end{equation}
\begin{equation}
\begin{split}
  \int^{u} 
  du^{\prime} 
  {\rm sn}^{4}(u^{\prime},k)
  =&
  \frac{{\rm sn}(u,k) {\rm cn}(u,k) {\rm dn}(u,k)}{3 k^{2}}
\\
  &
  +
  \frac{2+k^{2}}{3k^{4}}
  u
\\
  &-
  \frac{2(1+k^{2})}{3k^{4}}
  \mathsf{E}[{\rm am}(u,k),k],
\end{split}
\end{equation}
where $\mathsf{E}(u,k)$, ${\rm am}(u,k)$, 
and ${\rm dn}(u,k)$ are 
the second kind of incomplete elliptic integral, 
Jacobi amplitude function, 
and Jacobi elliptic function, respectively. 


\section{The effect of the field-like torque or Rashba torque}

The direction of the field-like torque or the Rashba torque is given by 
$\mathbf{m} \times \mathbf{p}$, 
where $\mathbf{p}$ is the direction of the spin polarization. 
This means that the effects of these torques can be regarded as a normalization of 
the field torque $\mathbf{m} \times \mathbf{H}$. 
Then, the energy density $E$ and the magnetic field $\mathbf{H}$ in the calculations of $\mathscr{W}_{\rm s}$ and $\mathscr{W}_{\alpha}$ 
should be replaced with 
an effective energy density $\mathcal{E}$ and an effective field $\bm{\mathcal{B}}$ given by 
\begin{equation}
  \mathcal{E}
  =
  E 
  - 
  \beta 
  M H_{\rm s}
  \mathbf{m}
  \cdot
  \mathbf{p},
  \label{eq:effective_energy}
\end{equation}
\begin{equation}
  \bm{\mathcal{B}}
  =
  \mathbf{H}
  +
  \beta 
  H_{\rm s}
  \mathbf{p},
  \label{eq:effective_field}
\end{equation}
where a dimensionless parameter $\beta$ characterizes the ratio 
of the field-like torque or Rashba torque to the spin Hall torque. 
We neglect higher order terms of the torque \cite{garello13} for simplicity, 
because these do not change the main discussion here. 
In principle, $j(E)$ satisfying Eq. (\ref{eq:condition_1}) can be obtained 
by a similar calculation shown in Sec. \ref{sec:Spin Hall system}. 
However, for example, 
the right-hand-side of Eq. (\ref{eq:jE}) now depends on the current 
through $\mathcal{E}$ and $\mathbf{H}$. 
Thus, Eq. (\ref{eq:jE}) should be solved self-consistently with respect to the current $j$, 
which is technically difficult. 



\begin{thebibliography}{50}
\bibitem[{com({\natexlab{a}})}]{comment1}
\bibinfo{note}{Equation (\ref{eq:jE_saddle}) is applicable to $H_{\rm t}>0$.
  When $H_{\rm t}>0$, the magnetization can move the curve except at a point
  $\mathbf{m}_{\rm saddle}$, and thus, $\mathscr{W}_{\rm s} \neq 0$, resulting
  in a finite $j(E_{\rm saddle})$. On the other hand, when $H_{\rm t}=0$, the
  magnetization cannot move at any point on the constant energy curve of
  $E_{\rm saddle}$, and thus, $\mathscr{W}_{\rm s}$ is zero.}
\end{thebibliography}




\end{document}